\documentclass[aps,twocolumn,floats,nofootinbib,groupedaddress]{revtex4-1}

\usepackage{amsmath,amssymb,amsfonts, graphics,graphicx,color}
\usepackage{float}

\newcommand{\ml}{\mathrm{ml}}
\newcommand{\mul}{\mathrm{\mu l}}
\newcommand{\mg}{\mathrm{mg}}
\newcommand{\Pa}{\mathrm{Pa}}
\newcommand{\pN}{\mathrm{pN}}
\newcommand{\mum}{\mu\mathrm{m}}
\newcommand{\nm}{\mathrm{nm}}
\newcommand{\rmd}{\mathrm{d}}

\begin{document}

\title{Collective force generation by groups of migrating bacteria}

\author{Benedikt~Sabass}
\author{Howard~A.~Stone}
\affiliation{Department of Mechanical and Aerospace Engineering, Princeton University, NJ 08544, USA}
\author{Joshua~W.~Shaevitz}
\email{shaevitz@princeton.edu}
\affiliation{Joseph Henry Laboratories of Physics and Lewis--Sigler Institute for Integrative Genomics, Princeton University, Princeton, NJ 08544}

\date{\today}

\begin{abstract}
From biofilm and colony formation in bacteria to wound healing and embryonic development in multicellular organisms, groups of living cells must often move collectively. While considerable study has probed the biophysical mechanisms of how eukaryotic cells generate forces
during migration, little such study has been devoted to bacteria, in particular with regard to the question of how bacteria generate and coordinate forces during collective motion. This question is addressed here for the first time using traction force microscopy. We study two distinct motility mechanisms of {\it Myxococcus xanthus}, namely twitching and gliding. For twitching, powered by type-IV pilus retraction, we find that individual cells exert local traction in small hotspots with forces on the order of 50~pN. Twitching of bacterial groups also produces traction hotspots, however with amplified forces around 100~pN. Although twitching groups migrate slowly as a whole, traction fluctuates rapidly on timescales <1.5 min. Gliding, the second motility mechanism, is driven by lateral transport of substrate adhesions. When cells are isolated, gliding produces low average traction on the order of 1~Pa. However, traction is amplified in groups by a factor of ~5. Since advancing protrusions of gliding cells push on average in the direction of motion, we infer a long-range compressive load sharing among sub-leading cells. Together, these results show that the forces generated during twitching and gliding have complementary characters and both forces are collectively amplified in groups.
\end{abstract}

\maketitle

\section{Introduction}

Many bacteria possess the ability to migrate over surfaces in large groups to facilitate such diverse phenomena as predation, aggregation, and biofilm formation. Research into the motility of microbes over the past few decades has made considerable progress towards an understanding of how single cells move, particularly the proteins involved, their regulation, and their ability to generate mechanical forces. However, the properties of the generated surface traction and the coordination of forces by multiple cells to produce coherent group motion remain unclear.

\begin{figure}
  \centering
	\includegraphics[width=\columnwidth]{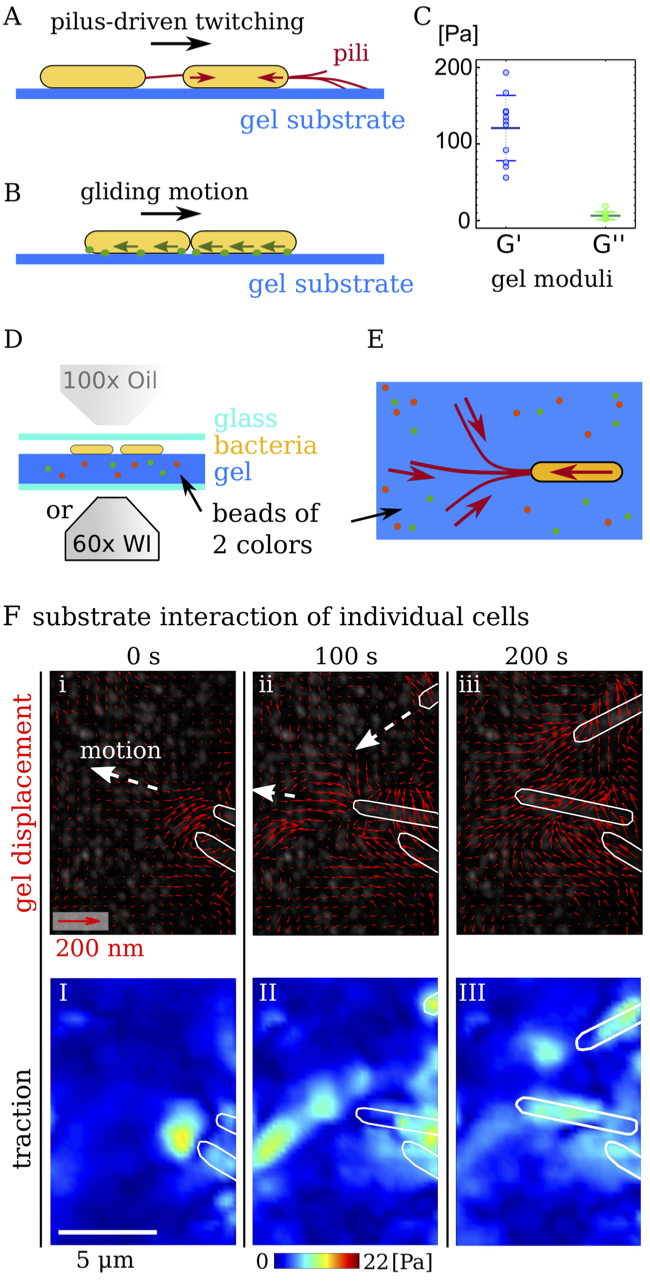} %
\caption{Measurement of substrate traction resulting from the two migration machineries of \textit{M.~xanthus}. A)~Type-IV pilus extension-retraction cycles allow bacterial motion referred to as twitching motility or S-motility. Pili can also mediate mechanical cell-cell coordination. B)~Gliding or A-motility results from lateral translation of transmembrane complexes along the bacterium. Steric interactions allow directional coordination of gliding cells. C)~The substrates employed for the experiments are polyacrylamide (PAA) gels with storage modulus G' and loss modulus G''. Mean $\pm$ standard deviation are plotted over measured data. E)~Sketch of the setup for traction force microscopy (TFM). Cells are placed on a gel containing fluorescent marker beads of two colors. Individual bacteria are imaged from above with a $100\times$ objective. For bacterial groups, a $60\times$ water immersion objective (WI) is used. E)~Top view of the setup. F)~Gel deformation and substrate traction resulting from motion of individual wild-type bacteria that can twitch and glide. White outlines show contours of bacteria. (i-iii)~Quiver plot of gel displacements. (I-III)~Calculated traction magnitude. } 
\label{fig_setup}
\end{figure}

\textit{Myxococcus xanthus} exhibits complex collective behaviors including vegetative swarming, predation, and fruiting body formation~\cite{keane2016predatory}. This organism is well-characterized and uniquely suited for motility studies. It employs two migration machineries~\cite{hodgkin1979genetics, zhang2012individual} to move in an intermittent forward-backward motion~\cite{welch2001cell, mignot2005regulated} (Fig.~\ref{fig_setup}A,B). First, twitching, sometimes called social (S), motility~\cite{mattick2002type, harshey2003bacterial, pelicic2008type, maier2015bacteria} is powered by the extension and retraction of type-IV pili, whereby extruded filaments adhere to the surface and filament retraction produces motility~\cite{wu1995genetic, wall1999type, merz2000pilus,skerker2001direct, marathe2014bacterial, chang2016architecture} (Fig.~\ref{fig_setup}A). Pili also mediate cell-cell adhesion and retraction has been shown to  be triggered by polysaccharides on neighboring cells~\cite{li2003extracellular, black2006type}. A second, genetically distinct, motility system~\cite{hodgkin1979genetics,youderian2003identification,nett2007chemistry} is termed gliding, or adventurous (A), motility. Here, a gliding transducer complex~\cite{luciano2011emergence} that spans the membranes and periplasm converts the transmembrane proton gradient into force~\cite{nan2011myxobacteria,sun2011motor}. Motion occurs through translation of substrate-adhesion sites along the cell body~\cite{mignot2007evidence, balagam2014myxococcus, nan2016novel} (Fig.~\ref{fig_setup}B). 

Although many of the molecular details of these two systems are known, it is unclear if individual cells produce any measurable force during migration, or if and how groups of cells coordinate these forces. Inertia  and hydrodynamic forces for these cells are negligible. For example, the drag force on a cell moving at a typical {\it Myxococcus} migration speed of $1\,\mum/\rm{min}$ is on the order of $10^{-2}\,\pN$. Large traction forces will only occur  if  cells need to overcome friction with the surface or if the translation machinery itself has internal friction, similar to the situation for eukaryotic cells~\cite{oliver1995traction,schwarz2013physics}. Collective migration of bacteria within a contiguous group is even less well understood. Could forces arise from a balance between cell-substrate and cell-cell interactions?
Would this balance be local, or span larger distances within the group? Might one have ``leader'' cells at the advancing front of the group that exert forces locally to pull along those cells in the ranks behind? Or do all cells move forward from the back and push the advancing group forward? Finally, what are the timescales of force reorganization in groups? These mechanical aspects of bacterial migration have to date remained largely inaccessible to direct experimental measurement. 

In this study, we report the first spatially resolved measurement of bacterial cell-substrate stress using Traction Force Microscopy (TFM)~\cite{dembo1999stresses, butler2002traction,sabass2008high, trepat2009physical}. We show that the forces produced by the two distinct migration machineries of \textit{M.~xanthus} have characteristic features. For individual bacteria, we find that pili produce measurable traction that is localized several micrometers ahead of the cell body. Thus, bacterial pili produce a dipolar traction pattern. During the collective motion of twitching groups, traction occurs in local hotspots that fluctuate on a timescale that is much shorter than the timescale of group migration. Gliding cells, on the other hand, show very low forces during gliding. However, once organized into dense groups where the bacteria are aligned, gliding produces measurable traction oriented preferentially in the direction of group motion and is distributed over large areas, indicating compressive load sharing among cells. 

\section{Results}
For traction measurements, we employ soft elastic substrates made from chitosan-coated polyacrylamide (PAA) with shear modulus $G' \simeq 121\,\Pa$. The substrates contain fluorescent marker beads of two colors, which increases the spatial resolution of TFM to about $0.5\,\mum$. As shown in Fig.~\ref{fig_setup}C,E, cells in suspension are placed on the gel and imaged from above or below. Lateral cell-substrate forces during cell migration produce deformation of the gel. Time-lapse imaging of the fluorescent beads allows measurement of a spatio-temporally varying deformation field relative to the first frame of a sequence. This deformation field is then used to calculate the relative traction that bacteria exert on to the substrate by making use of a regularized Fourier transform-based inversion technique~\cite{sabass2008high}. The traction calculated in this way is measured relative to the possibly pre-stressed first frame of an imaging sequence. To test if \textit{M. xanthus} produces any measurable substrate forces during migration, we investigated wild type cells with the ability to both twitch and glide. Figure~\ref{fig_setup}F shows representative results for the displacement field and traction maps that clearly demonstrate the presence of substrate forces below and ahead of migrating bacteria.

\subsection{Individual twitching cells produce small hotspots of traction}
\begin{figure}
  \centering
	\includegraphics[width=\linewidth]{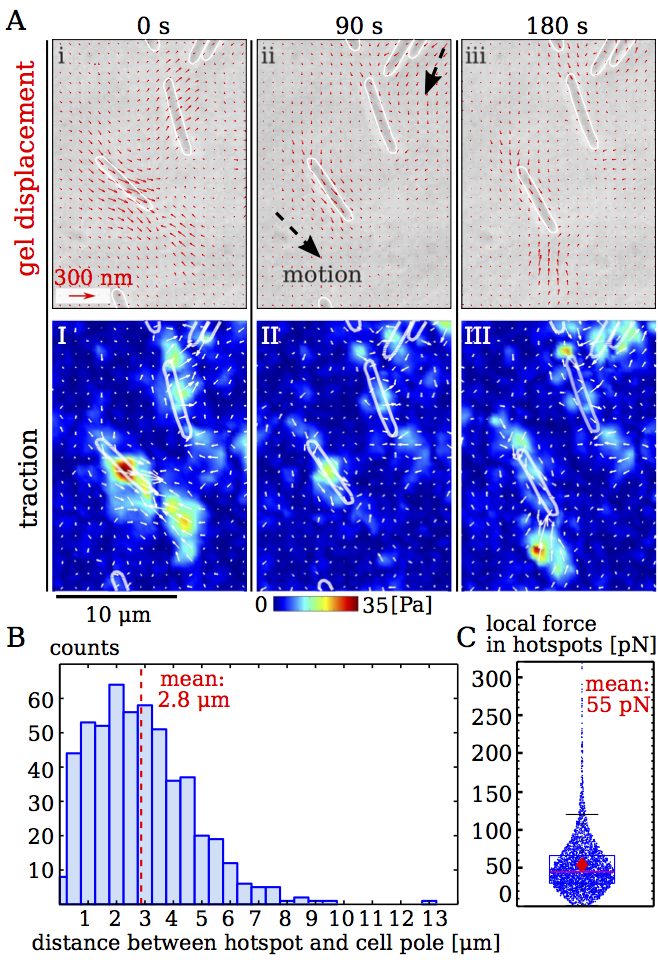} 
\caption{Twitching of individual, gliding-deficient bacteria that move by using their pili ($\Delta${\it aglQ}). A)~Gel deformation and substrate traction at three time points. White outlines show contours of bacteria. (i-iii)~Quiver plot of gel displacements. (I-III)~Calculated traction magnitude. Note that hotspots of appear in front of bacteria. B)~Distances between hotspots and the nearest cell pole. C)~Magnitude of overall force in individual hotspots as estimated by assuming point forces. Dots are individual measurements, box contains $[25-75]\%$ of data around the median, diamond shows the mean. Data for B and C was collected from $7$ experiments with $>5$ cells each.}
\label{fig_single_bacteria_pili}
\end{figure}

To isolate the different motility systems, we first probed twitching cells that lack the ability to glide due to a deletion of the {\it aglQ} gene. We observe localized areas of substrate deformation immediately in front of twitching cells, yielding bead displacements on the order of $100\,\nm$, see Fig.~\ref{fig_single_bacteria_pili}A(i-iii) and SI movie M1. The corresponding calculated traction is concentrated in hotspots, which have an apparent size on the order of $1\,\mum^2$ due to resolution limitations. Time-lapse images in Fig.~\ref{fig_single_bacteria_pili}A (I-III) demonstrate that the traction field is dynamic and changes on a timescale on the order of a minute. Among moving cells, not all show measurable traction at all times. If hotspots are present, we observe on average 2-3 of them, with as little as 1 and as many as 6. Hotspots in front of cells mostly do not stretch all the way to the cell bodies, which demonstrates that pili likely only engage the substrate at their tips. The distance between hotspots and the closest cell pole is on average $\sim 3~\mu\rm{m}$, but can be up to $14~\mu\rm{m}$, which is consistent with reported pilus lengths determined by electron microscopy~\cite{kaiser1979social} (Fig.~\ref{fig_single_bacteria_pili}B). 

Note that the long range of pili allows bacteria to connect to each other even when they are seemingly far apart. These invisible mechanical links among cells, together with resolution limitations, renders a detailed assessment of a force balance on the level of individual bacteria difficult. Also, the force applied at individual hotspots can not be estimated from local integration of the traction field since undersampling suppresses those high frequency spatial variations that affect the force magnitude strongest. Nevertheless, the clear localization of traction in hotspots makes it possible to estimate the overall force magnitude corresponding to each hotspot by assuming that forces are applied only at infinitesimal points at their center. For localized traction, this approach yields an improved estimate of force magnitude~\cite{sabass2008high}. Details of the method are described in the SI. For individual twitching bacteria, we find that the hotspots correspond on average to around $50\,\pN$, where almost all forces are smaller than $\sim 150\,\pN$ (Fig.~\ref{fig_single_bacteria_pili}C). These numbers may be compared with pilus retraction forces measured using optical tweezers~\cite{clausen2009high}. There, retraction of individual type IV-pili at \textit{M. xanthus} stalled at maximum forces of $\sim 149\,\pN$. For pilus retraction at speeds $\sim1\,\mum/s$, which roughly corresponds to the gliding speed of bacteria, $30\,\pN$ of force were measured. Thus, our first in-situ measurements of bacterial cell-substrate forces are consistent with results from other methods.

\subsection{Individual gliding cells exert very little traction}
\begin{figure}
  \centering
	\includegraphics[width=\linewidth]{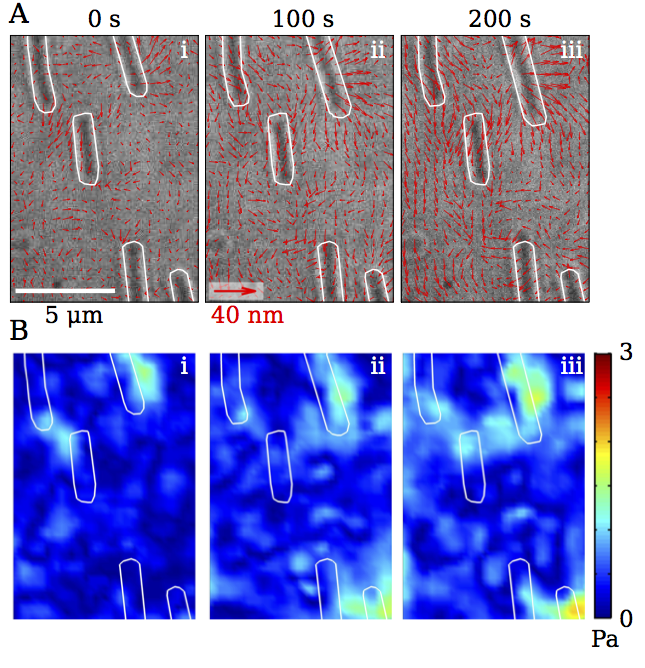} 
\caption{Gliding of individual bacteria from a strain without pili ($\Delta{\it PilA}$). A)~Quiver plots of gel displacement. B)~Calculated traction. The gel displacements beneath individual gliding bacteria are very small, $<20\,\nm$, and can hardly be distinguished from detection noise. As a result, traction estimation is error-prone.} 
\label{fig_single_bacteria_deltaPilA}
\end{figure}

To next investigate the motion of individual cells that do not employ pili but move by the complementary gliding mechanism, we
performed experiments using the twitching-deficient mutant $\Delta${\it pilA}. Fig.~\ref{fig_single_bacteria_deltaPilA} shows typical results for bacteria that move individually without contacting each other. Here, substrate deformation is below $20~\nm$ and very little overall traction is observed. Since displacements are close to the measurement precision, random noise is prominent. However, approximate co-localization of traction with bacteria implies that gliding bacteria do deform the substrate to some degree. Traction is not localized in front of the cells but beneath them. Estimated traction from individual gliding cells is on the order of $10$ times smaller than for twitching cells. Consequently, we conclude that gliding of individual cells is a low-friction process that hardly affects the environment mechanically.

\subsection{Groups of twitching cells exert local, fluctuating traction}
\begin{figure}
  \centering
	\includegraphics[width=\linewidth]{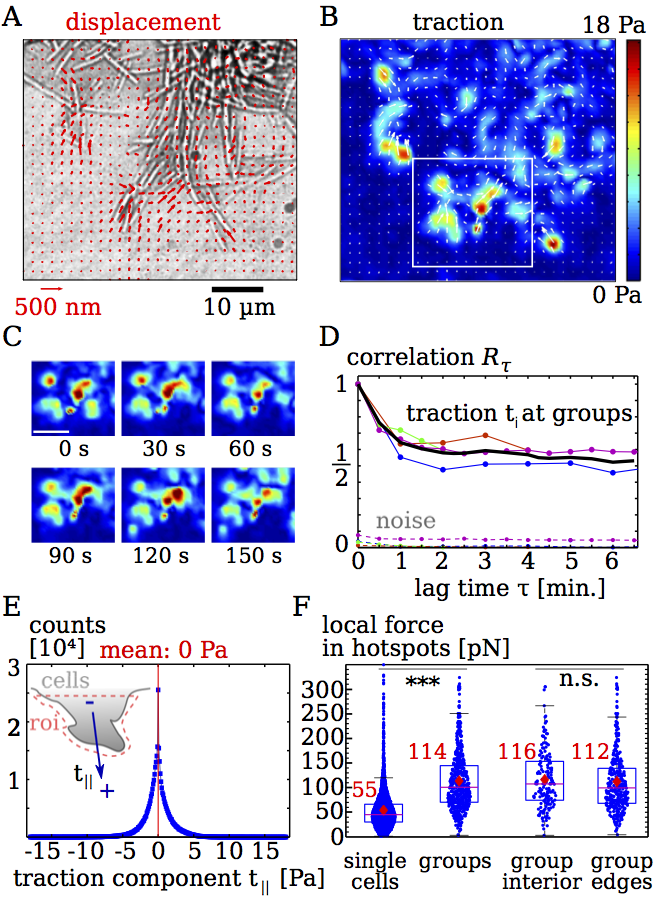} %
\caption{Collective migration of twitching bacteria that are gliding-deficient ($\Delta${\it aglQ}). A)~Gel displacements. Only every $4$th measurement is displayed for clearer visibility. B)~Calculated traction showing hotspots. C)~Snapshots displaying evolution of of the traction pattern inside the region denoted by a white rectangle in (B). Bar:~$10\,\mum$. D)~The autocorrelation $R_{\tau}$ quantifies temporal fluctuations of traction. Upper lines: data below cell groups and mean (black line). Lower, dotted lines: traction noise measured in regions without cells. E)~Distribution of the traction component $t_{||}$ resulting from projection of $t_i$ on the average orientation of a protruding group in a region of interest (roi). Tractions have a vanishing mean, showing that cell-substrate forces balance locally. Bin width $0.1\,\Pa$. F)~Force magnitude of hotspots. Single twitching cells produce significantly weaker hotspots than groups. In groups, forces at the edge are comparable to those measured below the cells. Dots are individual measurements. Data for cell groups in D),E) collected from $4$ separate experiments with an overall of $67$ images taken at frame rates of $[30-60]\,s$.
} 
\label{fig_group_bacteria_pili}
\end{figure}

To investigate how  groups of twitching cells distribute force while performing collective motility, we examined groups of twitching $\Delta${\it aglQ} cells~\cite{sun2011motor}. Gliding deficient mutants form slightly disorganized groups, where individual cells are not strongly aligned with each other. When deposited on a substrate, initial clumps of bacteria tend to spread out only slightly during the observation time. TFM analysis (Fig.~\ref{fig_group_bacteria_pili}A,B) shows highly localized substrate forces in spots at the periphery of the group. The traction from the outermost spots points towards the cell group, as is expected from a pulling action of the pili. The snapshots of traction magnitude shown in Figure~\ref{fig_group_bacteria_pili}C demonstrate that the localization and magnitude of forces is quite dynamic. To assess the traction dynamics quantitatively, we employ a correlation measure $R_{\tau}$ based on the definition
\begin{align}
R'_{\tau} = \frac{2}{N\,M} \sum_{n=1}^{N/2}\sum_{m=1}^{M}\left[\tilde{t}_{x,m,n} \tilde{t}_{x,m,n+\tau/\Delta} + \tilde{t}_{y,m,n} \tilde{t}_{y,m,n+\tau/\Delta}\right].
\end{align} 
Here, $\tilde{t}_{(x,y),m,n}$ is the mean-subtracted $(x,y)$-component of traction at position $m \in [1,\ldots,M]$ in the movie frame $n \in [1,\ldots,N]$. $\Delta$ denotes the time between each frame. The lag time $\tau$ of the correlation is in the range $[0,\Delta N/2]$. The traction correlations are measured in the vicinity of the cells. As a reference, we also record correlations of traction far away from cells. Then, correlations of real traction and noise are both normalized by the zero-lag correlation of real traction $R'_{0}|_{\rm{cells}}$ as $R_{\tau} \equiv R'_{\tau}/R'_{0}|_{\mathrm{cells}}$. 

For long times, $R_{\tau}$ approaches a non-zero constant $\simeq 0.5$, which results from measuring traction with respect to a prestressed state (see SI). In Fig.~\ref{fig_group_bacteria_pili}D, correlation data from four different experiments is shown where we distinguish between traction beneath cell groups and traction noise occurring away from cells. The noise correlation is clearly much smaller than the real signal. The correlations of traction decays very rapidly on timescales $\sim 1\,\rm{min.}$, which demonstrates the presence of rapidly fluctuating traction. These fluctuations can also be observed in the raw data images, where rapid, local displacements of marker beads occur (SI movie M2a,b). The movie also demonstrates that individual cells move rapidly in a seemingly random fashion, while the group edge expands, but only marginally.

We next assess the orientational ordering of forces in the protruding edge of groups to see if the leading cells are either pulling the group forward or being pushed by the group. As shown in the inset of Fig.~\ref{fig_group_bacteria_pili}E, we manually select regions of interest around groups of protruding cells and record the traction components $t_{||}$ that are aligned with the protrusion direction. Positive values of $t_{||}$ correspond to forces that push in the direction of the protrusion. The measured distribution of $t_{||}$ is symmetric with a center and mean value of $0\,\Pa$. On average, no pushing or pulling force occurs in protrusions. but rather forces balance locally. Thus, we conclude that forces are balanced in a local tug-of-war among twitching bacteria in protruding groups, with edge cells providing the traction that powers expansion.

Typical numbers of pili per \textit{M.~xanthus} bacterium have been reported to be around 4--10~\cite{kaiser1979social}, where in some cells up to $50$ pili were observed. Given the large number of potentially active pili in groups, it is not obvious that forces should be concentrated in the observed hotspots. However, if concentrated, the large number of available pili could produce strong forces on the order of nN, which would be comparable to those produced by much larger eukaryotes~\cite{balaban2001force,tan2003cells}.
Moreover, engaging the substrate with many pili simultaneously could potentially lead to very slow dynamics since motion would require
detachment of many pili. To clarify this issue, we compare the absolute force magnitude of traction hotspots at groups with the magnitude of hotspots at individual cells (Figure~\ref{fig_group_bacteria_pili}F). While hotspots at individual cells have a magnitude $\sim 50\,\pN$, we find for hotspots at groups a mean force of $114\,\pN$ with an uncertainty approximately as large as the mean. Thus, forces are amplified in groups by about a factor of two. Since these $\sim 100\,\pN$ are smaller than the maximum stall force of $\sim149\,\pN$~\cite{clausen2009high}, our result are still compatible with the notion that each traction hotspot in groups is caused only by one or a few pili. Furthermore, if many pili would cooperate to produce one traction hotspot, weaker forces would be expected for the edge of groups where fewer pili are present. However, a comparison of force magnitude shows no significant differences for hotspots at group edges or below the group interior. Together, we find that while groups of \textit{M.~xanthus} likely only use relatively few pili to simultaneously produce substrate forces, forces are nevertheless considerably amplified in groups when compared to single cells.

\begin{figure}
  \centering
	\includegraphics[width=.9\linewidth]{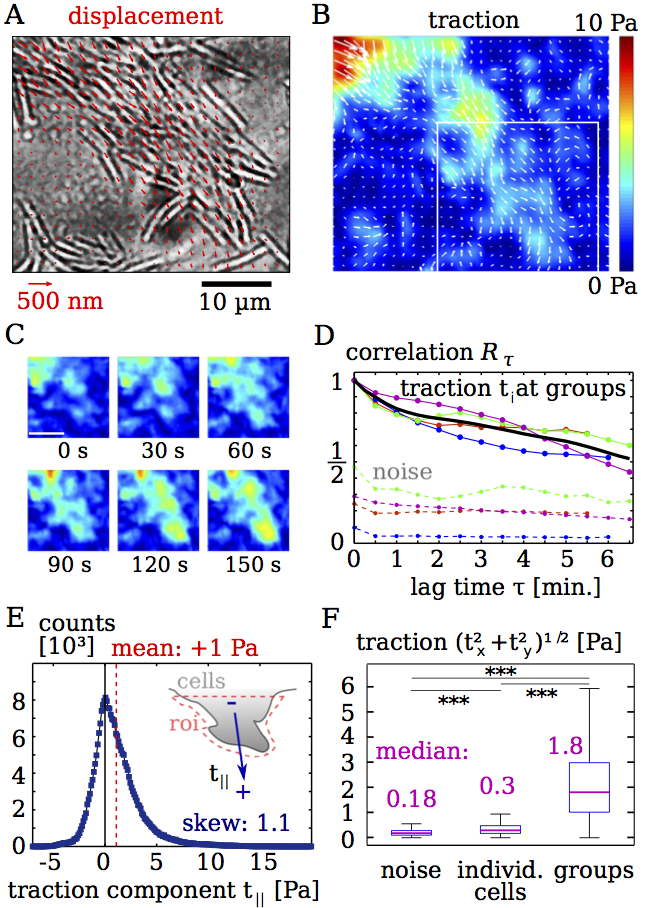} %
\caption{Collective migration of twitching-deficient strains ($\Delta${\it PilA}). A)~Groups move with a finger-forming spreading pattern. Red quivers are gel displacements. Only every $4$th measurement is displayed for clearer visibility. B)~Calculated traction. Note that the traction points on average in the direction of motion. C)~Snapshots displaying evolution of of the traction pattern inside the region denoted by a white rectangle in (B). Bar:~$10\,\mum$. D) Traction autocorrelation $R_{\tau}$. Upper lines: data from $4$ cell groups with mean (black line). Lower, dotted lines: traction noise measured in regions without cells. E)~Distribution of the traction component resulting from projection onto the average orientation of a fingering structure. Histogram bin width $0.126\,\Pa$. Data recorded from areas where bacteria form a fingering structure. Positive values of the mean traction $\langle t_{||}\rangle \simeq 1.1\,\Pa$ and the skewness demonstrates that twitching-deficient mutants in a growing finger tend to push the substrate in the direction of motion. F)~Bar plots of traction magnitude $\|t\|$ comparing noise away from cells, individual gliding $\Delta${\it PilA} cells, and compact groups of $\Delta${\it PilA} cells. Data for individual cells recorded from overall $28$ cells in $4$ experiments. Data for cell groups in D),E) are collected from $4$ experiments with each $\sim 25$ image frames. } 
\label{fig_deltapilA_group}
\end{figure}

\subsection{Gliding groups can exert persistent, coordinated force}
To next assess the collective mechanics of bacterial gliding, we probed groups of gliding $\Delta${\it pilA} cells. When placed on the imaging substrate, clumps of bacteria present at the start of the experiment spread in a fingering fashion, where the fingers consist of closely packed bacteria that move parallel to each other. We find that although gliding of individual cells does not produce much traction, gliding motion in groups leads to measurable forces (Figs.~\ref{fig_deltapilA_group}A,B). Here, traction is distributed in diffuse patches underneath the moving group and the traction magnitude is lower than in the presence of pili. Furthermore, the cell-substrate traction in the shown protrusion appears rather coordinated since the traction points in the direction of the advancing cells. 

The snapshots of traction magnitude shown in Fig.~\ref{fig_deltapilA_group}C illustrate that traction is dynamic, but changes appear less abrupt than for twitching cells. To assess the traction dynamics quantitatively, we calculate the correlation measure $R_{\tau}$ for the twitching-deficient mutants (Fig.~\ref{fig_deltapilA_group}C). Again, the correlation measure is normalized by the zero-lag autocorrelation below the cell groups in each movie $R'_{0}|_{\rm{cells}}$. The contribution of measurement noise is here evidently stronger than in the case of twitching motion due to the lower force magnitude. We find that the traction correlation does not show a rapid decay on short timescales as in Fig.~\ref{fig_group_bacteria_pili}D; instead, it decays over many minutes (at least as long as the duration of our experiment). Thus, gliding of groups causes traction variations that are slower than those resulting from pili.

Since the traction images of gliding groups in Fig.~\ref{fig_deltapilA_group}B suggest a ``pushing'' nature of the forces under advancing fingers, we quantitatively assess the directionality of forces in Fig.~\ref{fig_deltapilA_group}E. We manually select regions of interest around protruding fingers and record the traction components $t_{||}$ that are aligned with the protrusion direction. In contrast to the results from twitching cells, the distribution of $t_{||}$ is here asymmetric and heavy on the positive side, as quantified by a positive skew of $1.1$. The pushing nature of cell-substrate traction below advancing fingers is corroborated by a positive distribution median of $+0.62\,\Pa$, where the hypothesis of a vanishing median is rejected with $\simeq100\%$ confidence by a sign test. These pushing forces necessarily require long-range load balance, where compression of cells at the rear end of the protruding finger balances pushing forces at the tip (see also SI).

Finally, we  compare the forces produced by gliding groups with forces from individuals cells that are not touching each other. Since traction from gliding is distributed below the bacteria, we can not use the assumption of discrete point forces to calculate absolute force values. Instead, we record the distribution of traction magnitude $\|t_m\| \equiv \sqrt{t_{x,m}^2 +t_{y,m}^2}$ at every position $m$ either directly beneath individual cells or beneath densely packed groups. To obtain an estimate of the noise magnitude, traction magnitudes in areas without cells are also recorded. We find that individual gliding cells exert traction that is significantly above the noise threshold, but nevertheless quite weak with a median below $1\,\Pa$. Unexpectedly, we found that groups of gliding cells cells produce much higher traction than individual cells. The median of traction measured below gliding groups is more than 5 times higher than the median traction below individual cells, as shown in Figure~\ref{fig_deltapilA_group}F.

\section{Discussion}
In this study, we perform the first spatially resolved analysis of traction exerted by bacteria. We present definitive evidence for two very distinct patterns of force organization during the migration of {\it Myxoccocus xanthus}. 

In particular, pilus-driven twitching of individual cells can lead to a tug-of-war like motion where bacteria exert counteracting forces on the substrate. Here, we find cell-substrate traction that is concentrated in hotspots with a force magnitude on the order of $\sim 50\,\pN$. When the bacteria form a group, the number of available pili per substrate area is increased. Therefore, one might expect that twitching groups do not produce traction hotspots, but instead a rather continuous traction pattern with coordinated directionality.  However, we observe that groups exhibit similar hotspots of traction as individual cells, albeit with a significantly amplified force magnitude around $100\,\pN$. Possible explanations for the collective force amplification include biochemical regulation~\cite{nudleman2005cell,konovalova2011close}, cellular jamming leading to higher resistance and thereby higher force generation in the retraction motors~\cite{clausen2009high}, and the collective action of pili. Traction hotspots are rather short-lived and decay typically within the minute timescale. Moreover, pilus forces in groups are not coordinated, such that the groups as a whole hardly move over the timecourse of $10-20$~min. Overall, we conclude that the force from pili is not employed efficiently for the purpose of migration. However, pili clearly provide mechanical anchoring to the substrate and one might speculate about a potential sensory role that pilus retraction plays in allowing cells to probe their mechanical surroundings~\cite{persat2015type}.

For gliding cells, we find almost the opposite results. Gliding is currently thought to be powered by elastically connected adhesion sites that are stationary with respect to the substrate~\cite{balagam2014myxococcus}. Once these adhesion sites reach the rear pole, the machinery is disassembled~\cite{treuner2015small}. If this adhesion disassembly can not keep up with the migration speed, one would expect traction at the lagging pole. Such traction was indeed found for an unrelated gliding of apicomplexans~\cite{munter2009plasmodium}. However, we did not find pronounced traction at the rear end for $M.~xanthus$, suggesting that gliding is not a slip-stick motion limited by mechanical adhesion detachment in the studied conditions. In spite of the low forces measured for individual gliding cells, groups of gliding bacteria exert measurable substrate traction that can push in the direction of motion. To balance this traction, cells must experience long-ranged compressive force. While contact-dependent biochemical mechanisms can affect gliding~\cite{jakobczak2015contact}, traction under gliding groups may also originate from a mechanical cell-cell interaction since random motion reversal of individual cells leads to stalling forces on other cells~\cite{zhang2011quantifying}. In this picture, velocity variation produces an innately integrative mechanism for maintaining directional load while allowing group rearrangement.

On a technical level, our study is limited by the minuscule size of bacteria and pili. The spatial resolution of TFM results is limited by the density of measurements of the substrate deformation. Using fluorescent beads of two colors, substrate deformation can be measured approximately every $0.5\,\mum$, which is comparable to the bacterial thickness. Therefore, we expect that the real traction exerted by bacteria varies on a lengthscale comparable to, or shorter than our measurement scale. We are then dealing with a spatially undersampled traction field, which is a problem that is routinely encountered in the context of TFM at eukaryotic adhesion sites. A consequence of undersampling of the displacement field is that the absolute traction magnitude is usually underestimated~\cite{sabass2008high, stricker2010optimization}.

While we approach the current spatio-temporal limits of TFM in this study, we are still able to compare relative load values and to assess the spatiotemporal organization of traction. A number of challenging refinements of the methodology are desirable. First, accurate three-dimensional tracking of beads in the substrate would possibly allow the assessment of vertical forces and allow for precise determination of the vertical distance between the beads and the bacteria. Such analysis was  precluded in our studies by bacterial photodamage from the fluorescence excitation light. Second, the gel displacements are measured with respect to a prestressed state since it proved difficult to recover the fully relaxed state after removal of bacteria. Although not essential for this study, it is generally desirable to obtain the relaxed state of the substrate. Third, comparison of TFM results with standard bacterial migration assays would be facilitated by the use of agar-based TFM substrates. Preliminary tests indicated that \textit{M.~xanthus} does not deform agar appreciably, which is likely a result of the larger rigidity as compared to PAA.

Many facets of bacterial cell-surface interactions are yet poorly understood~\cite{taktikos2013motility, persat2015mechanical}. For example, it remains to be explained why twitching \textit{M. xanthus} moves faster on soft agar than on stiff agar~\cite{shi1993two}. Controlled, biochemical responses to force occur, inter alia, during surface-dependent virulence of \textit{P. aeruginosa}~\cite{siryaporn2014surface,persat2015type} or inside epithelial host cells in contact with \textit{N. gonorrhoeae}~\cite{higashi2007dynamics}. We have shown that the combination of traction measurement with genetic or biochemical perturbations provides a viable and fruitful approach to improve our understanding of bacterial mechanics and address these fundamental questions.

\section{Acknowledgments}

We acknowledge invaluable advice and help with experimental protocols from S.~Thutupalli and G.~Liu.
We also acknowledge help from G.~Laevsky at the Confocal Microscopy Facility of Princeton University. A.~Perazzo 
is thanked for advice concerning rheometry. T\^{a}m Mignot is thanked for providing bacterial strains.
This work was supported by NSF award PHY-1401506, and B.S. was supported by the NSF award MCB-1330288 (to H.A.S. and Z.~Gitai) and by a fellowship from the German Academic Exchange Service (DAAD).

\section{Methods}

\subsection{Cell culture}
\textit{M. xanthus} strains employed in this study are a wild-type DZ2 strain, the gliding-deficient strain TM 146 DZ2 $\Delta${\it aglQ} \cite{sun2011motor}, and a pilus-deficient strain DZ2 AglZ-YFP $\Delta${\it pilA}. Bacteria are grown overnight at $32^{\circ}\,\mathrm{C}$ in CYE medium at $\mathrm{pH}~7.8$ consisting of $1\,\%$ (w/v) Casitone, $0.5\,\%$ yeast extract, 10 mM 3-(N-morpholino) propanesulfonic acid (MOPS), and 4 mM~$\mathrm{MgSO}_4$~\cite{bustamante2004analysis}. To remove nutrients from the medium prior to experiments, cells are washed once in TPM (10 mM Tris-HCl, pH 7.6, 1 mM~$\mathrm{KH}_2\mathrm{PO}_4$, 8 mM~$\mathrm{MgSO}_4$). The suspension of bacteria in TPM is briefly vortexed to ensure homogeneity before depositing it on the gel for imaging.

\subsection{Preparation and characterization of elastic substrates for TFM}
Polyacrylamide gels were prepared as described in~\cite{sabass2008high} for use with fluorescent beads of two colors. We 
prepare $250\,\mul$ gel with final concentrations of $3\%$~polyacrylamide and $0.06\%$~ bisacrylamide. The gel consists of water, polyacrylamide solution ($40\,\%$), bisacrylamide solution ($2\,\%$), each $4.5\,\mul$ of orange and dark red fluorescent beads (FluoSpheres, diameter $0.040\,\mum$, carboxylate-modified, (565/580)~nm and (660/680)~nm), and $1.5\,\mul$ of freshly prepared ammonium persulfate solution ($10\,\%$ in water). Polymerization is initiated with $0.75\,\mul$ of N,N,N',N'-tetramethylethylenediamine (TEMED). $40\,\mul$ of the forming gel is spotted on a plasma-treated microscope slide or a glass-bottom petri dish and covered with a hydrophobic cover slip. After waiting for one hour to let the gel polymerize, the top coverslip is carefully removed and the gels are washed with water. If washing of gels is insufficient, bacteria can not survive on the gel, which we attribute to unpolymerized gel constituents.

Elastic properties of the PAA gel are measured by a stress-controlled rheometer (Anton Paar, Physica MCR 301). All of the rheometry measurements are carried out at $23^{\circ}\mathrm{C}$. To avoid slippage between the gel and rheometer, we employ a parallel-plate geometry with sand-blasted plates of roughness $[8-9]\,\mum$ (PP50/S). The gap thickness is chosen to be $0.5\,\rm{mm}$. Using a cone-plate geometry to obtain a homogeneous velocity gradient throughout the sample (Measuring cone CP50-1/TG) yielded similar results. After preparing the gel and adding the polymerization initiator, the liquid is placed on the rheometer and the measurement plate is moved into measurement position. After letting the gel polymerize between the plates for $15\,\rm{min}$, the rim of the plates is covered with a small film of water to avoid evaporation. Polymerization is allowed to proceed for $45-60\,\rm{min}$ before data recording to ensure that the elastic properties reached stationary values. Gelation of the substrate produces significant normal forces on the rheometer plate, which can affect the measurement of shear moduli. Therefore, the normal forces are set to zero by slight adjustment of the gap size before before commencing the measurement. For the employed PAA gel ($3\%$~PAA, $0.06\%$~BIS), we obtain a shear modulus of $G' = 121\,\pm 43\,\Pa$ ($11$ gels measured) (see Ref.~\cite{flanagan2002neurite} for literature values). Assuming a Poisson ratio close to $1/2$~\cite{boudou2009nonlinear}, the Young modulus is estimated as $E = 2(1 + \nu) G'\simeq 360\,\Pa$. At typical oscillation frequencies of $[0.01\ldots 10]\,\mathrm{Hz}$, the loss modulus is found to be small, $G''=6\,\pm5\,\Pa$.

\subsection{Coating of substrates with chitosan}
Chitosan, a deacetylated form of chitin, is a
polysaccharide that has broad biocompatibility. We
found that PAA gels coated with chitosan can be used
as motility assays where, depending on the
concentration of chitosan, myxobacteria can move individually and in groups~\cite{ducret2012wet}. 
To coat the gels, we dissolve $10\,\mg$ chitosan in $3\,\ml$ of $0.2\,\mathrm{M}$ acidic acid by gentle pipetting. 
The solution is then diluted $1/50$ with DI-water. After gently removing excess water from the gel surface, $100\,\mul$ 
of the chitosan solution is placed on the gels and left there for at least one hour. Prior to imaging, gels are washed 
three times with a tris buffer solution. To prepare the sample for imaging, about $7\,\mu\mathrm{l}$ of cell suspension 
in TPM are spotted on the gel and excess liquid is removed with a tissue. Finally, a cover slip is gently placed on top of the 
sample.

\subsection{Imaging}
Imaging is done on a Nikon Ti-E confocal microscope with Perfect Focus System, where a Yokogawa spinning disc (CSU-21) is mounted with a quad dichroic accommodating lasers with wavelengths of $405$, $488$, $561$, and $647\,\nm$. Images are taken with a Hamamatsu ImageM back thinned EMCCD or an ORCA Flash digital CMOS camera. Individual bacteria are imaged through the glass coverslip above the cells using a $100\times$ oil immersion objective with $1.5$ magnification. To avoid applying vertical pressure on bacterial groups that are occasionally
thicker than the slit between sample and gel, we image groups from below by using a glass-bottom dish as support for the gel. Focusing through the whole gel then requires matching the refractive index by use of a $60\times$ water immersion objective with $1.5$ magnification.

\subsection{Traction reconstruction}
To avoid evaporation and allow bacterial migration, it was necessary to cover the sample with a glass 
slip, where a thin spacer maintained a micron-scale distance between the glass slip and gel. If one wishes
to image only the gel without bacteria to obtain a stress-free reference image of the beads, the delicate 
setup would have to be disassembled under the microscope. This task proved unfeasible. Therefore, a stress-free 
reference image for tracking of the fluorescent marker beads is not available. Thus, we employed the first frame of a 
time-lapse series as reference frame. Consequently, the displacements and calculated tractions are those that occur relative 
to the first frame of a movie. Computational analysis of gel deformation and traction force estimation is done as described 
previously~\cite{sabass2008high, plotnikov2014high}. Briefly, we employ a correlation-based
tracking procedure that allows to extract deformation information simultaneously from both 
image channels. At the position $x,y$, the displacement field $u_i(x,y)$ is related to a 
traction field $t_i(x',y')$ through convolution with a Green's function $G_{ij}$ as
$u_i(x,y)= \int G_{ij}(x-x',y-y')t_i(x',y')\,\rmd x' \rmd y'$ where the integral extends over the whole
gel surface plane~\cite{landau1986course}. We estimate the traction field that produces the measured displacement field by inverting 
the convolution equation in Fourier space while regularizing the traction
magnitude. A constant regularization parameter value was employed for the whole data analysis. 
Edge-effects resulting from solving the system in Fourier space are avoided by zero-patterning 
the edges of the displacement field. For inferring the force magnitude at hotspots by assuming point forces, we employ the established procedure of matrix inversion using a singular value decomposition and Tikhonov regularization~\cite{schwarz2002calculation,sabass2008high}. As an important improvement, we assume here that the beads are located a finite distance beneath the surface of the gel (see SI). Histograms in figures were prepared by choosing a bin width $w$ that is close to the value given by the heuristic rule $w = 3.49\,\sigma\,\mathcal{N}^{-1/3}$, where $\sigma$ is the standard deviation of the data and $\mathcal{N}$ is the number of data points~\cite{scott1979optimal}. Boxplots in the presented figures show the $25\%-75\%$ range of the distributions around the indicated medians. The significance of having two different distributions is quantified with a rank sum test.

\section{Appendix}
\subsection{Estimation of point force magnitude}
\begin{figure}[H]
  \centering
	\includegraphics[width=.9\linewidth]{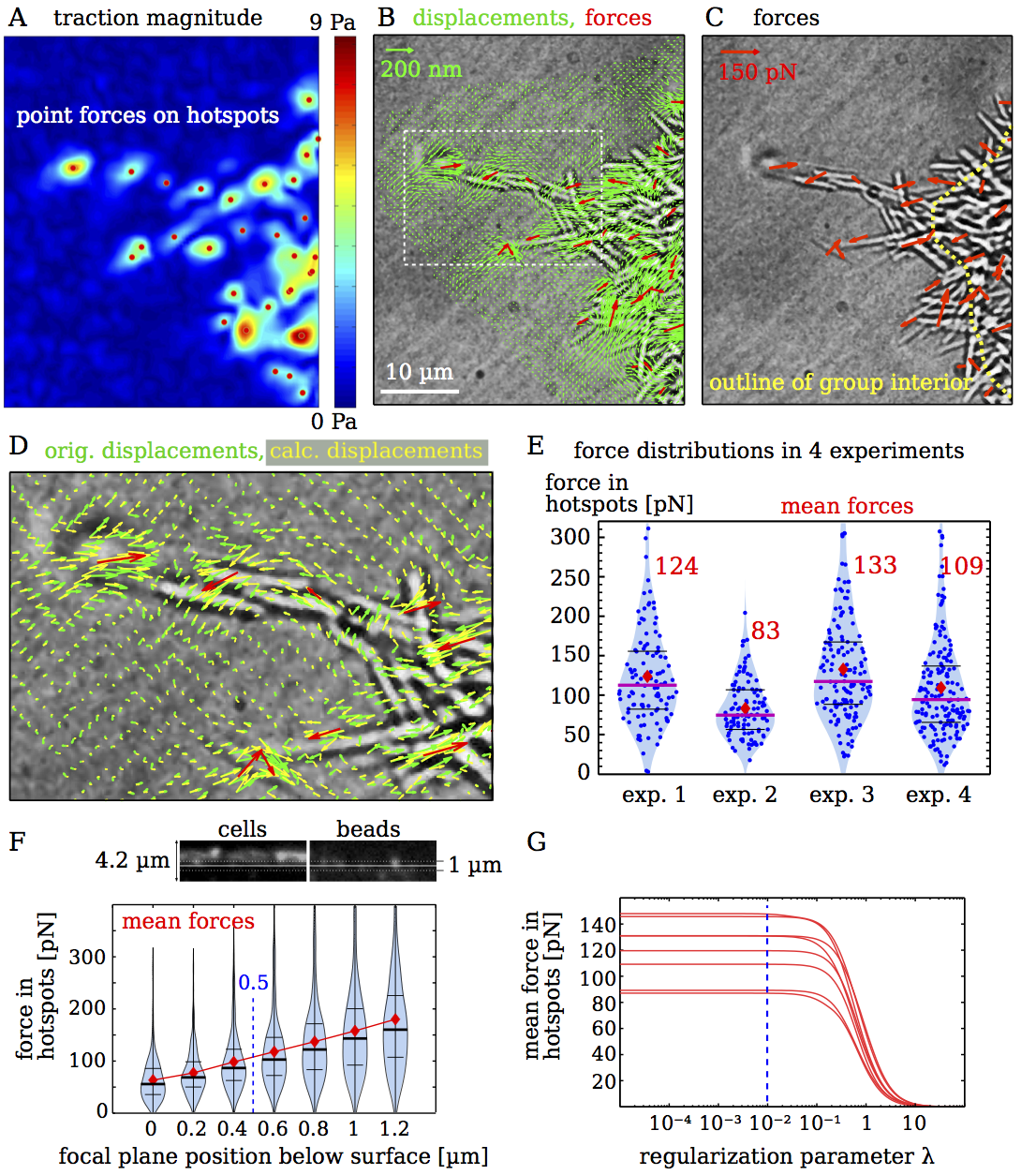} %
\caption{Estimation of force magnitude in traction hotspots created by pili. A)~Reconstructed traction field contains hotspots. Using these traction maps, a point is placed inside each hotspot. B)~Assuming that all force is concentrated at these points, we perform a maximum likelihood estimate of forces from the measured displacements. C)~Point forces form rather disorganized patterns with opposing forces being close to each other. D)~Magnified displacement data from the region of interest indicated in (B). Green quivers show original displacements, yellow quivers show the displacements that were back-calculated from the inferred point-forces. For clearer visibility, only every second quiver is shown. E)~Comparison of force magnitudes measured in 4 separate experiments with each 4 images. Data in (A-D) is from experiment 2. F)~Dependence of force on vertical position of the imaging plane below the gel surface. Due to the finite point spread function, beads lie slightly below the surface of the gel. Images show side-view of cells with beads and the position of the focal plane. The force magnitude increases with the assumed vertical distance between the gel surface and the plane in which displacements are measured. A distance of $0.5\,\mum$ is consistent with the vertical image scans and was therefore used for the analysis. F)~Dependence of average point forces on the regularization parameter $\lambda$ for different experiments. $\lambda$ is given in units of $1/\mathrm{pix}$ since displacements are measured in pixels and forces are scaled by $\Pa\,\mum^2/\mathrm{pix}^2$. We employ $\lambda = 0.01$ to regularize force magnitude as little as possible and consistently for all data sets.
} 
\label{fig_SI3_TRPF}
\end{figure}

For inferring the force magnitude at hotspots, we modify established methods~\cite{schwarz2002calculation,sabass2008high}. We assume Cartesian coordinates $x_1,x_2,z$, where an elastic material is bounded by the $x_1,x_2$ plane and occupies the upper half space $z>0$. The elastic Young modulus is denoted by $E$ and Poisson's ratio is denoted by $\nu$. Forces acting in the $x_1,\,x_2$ plane are denoted by $F_{1,2}$. Vertical forces are assumed to be zero. The displacements in a plane parallel to the material surface are denoted by $u_i(x_1,x_2,z)$, where we assume that $z\geq0$ is a constant. In a linear framework, we first consider a point force applied at the origin. Then, material displacements are related to the force through $u_i(x_1,x_2,z) = \sum_{j=1,2}G_{ij}(x_1,x_2,z) F_j$ with a Green's function~\cite{landau1986course}
\begin{align}
\begin{split}
&G_{ij}(x_1,x_2,z) =  \\
&\frac{(1+\nu)}{2\pi E}\left[\frac{2(1-\nu)r+z}{r(r +z)}\delta_{ij} + \frac{(2 r\,(\nu\,r+z)+z^2)\,x_i x_j}{r^3(r + z)^2}\right].
\end{split}
\end{align}
The measured displacements can be treated as resulting from the superposition of point forces at various locations. Thus, we can write a linear system relating displacements with index $n$ to point forces with index $m$ as $u_{i,n} = \tilde{G}_{i j, n m} F_{j,m}$. We employ an established procedure to solve the inverse problem by using a singular value decomposition of $\tilde{G}_{i j, n m}$ and Tikhonov regularization where the expression $\sum_{i,j,n,m}|u_{i,n} - \tilde{G}_{i j, n m} F_{j,m}|^2+\lambda^2|F_{j,m}|^2$ is minimized. Here, $\lambda$ is a regularization parameter. Figures~\ref{fig_SI3_TRPF}A,B,C illustrate the process of force estimation. First, we employ the results from standard traction force microscopy to localize hotspots of traction. Then, we manually place a point force into the center of every hotspot. Using the such defined point force locations together with the measured substrate displacements, we calculate the forces $F_i$. In Fig.~\ref{fig_SI3_TRPF}D we plot the originally measured displacements together with the displacements resulting from the calculated forces. Differences between measurements and calculation illustrate the approximative nature of the technique. In Fig.~\ref{fig_SI3_TRPF}E we compare the results from four independent experiments. Average values of forces differ slightly, but are on the order of $100\,\pN$. In Fig.~\ref{fig_SI3_TRPF}E we investigate the influence of the vertical distance $z$ between imaging plane and gel surface. Side-view of a vertical scan through the sample illustrates our choice for the imaging plane. As a result of the finite point spread function and the difficulty to determine the exact location of the gel surface, the value of $z$ is uncertain. However, we determined $z$ to lie in the range of $[0.2-1]\,\mum$ and therefore assume a constant value of $z=0.5\,\mum$ for all experiments. From Fig.~\ref{fig_SI3_TRPF}E, it can be seen that the force magnitude varies depending on the choice of $z$. Therefore, incorporating the position of the focal plane into the calculation results in a significantly improved force estimate. Fig.~\ref{fig_SI3_TRPF}F shows our choice of the regularization parameter. The magnitude of forces decreases rather sharply at $\lambda \sim 1$. To maintain consistency among the samples, we employ for all experiments $\lambda =0.01$. This value is for all samples below the transition to the regularization-dominated regime where force magnitude is strongly suppressed.

\begin{figure}[H]
  \centering
\includegraphics[width=.9\linewidth]{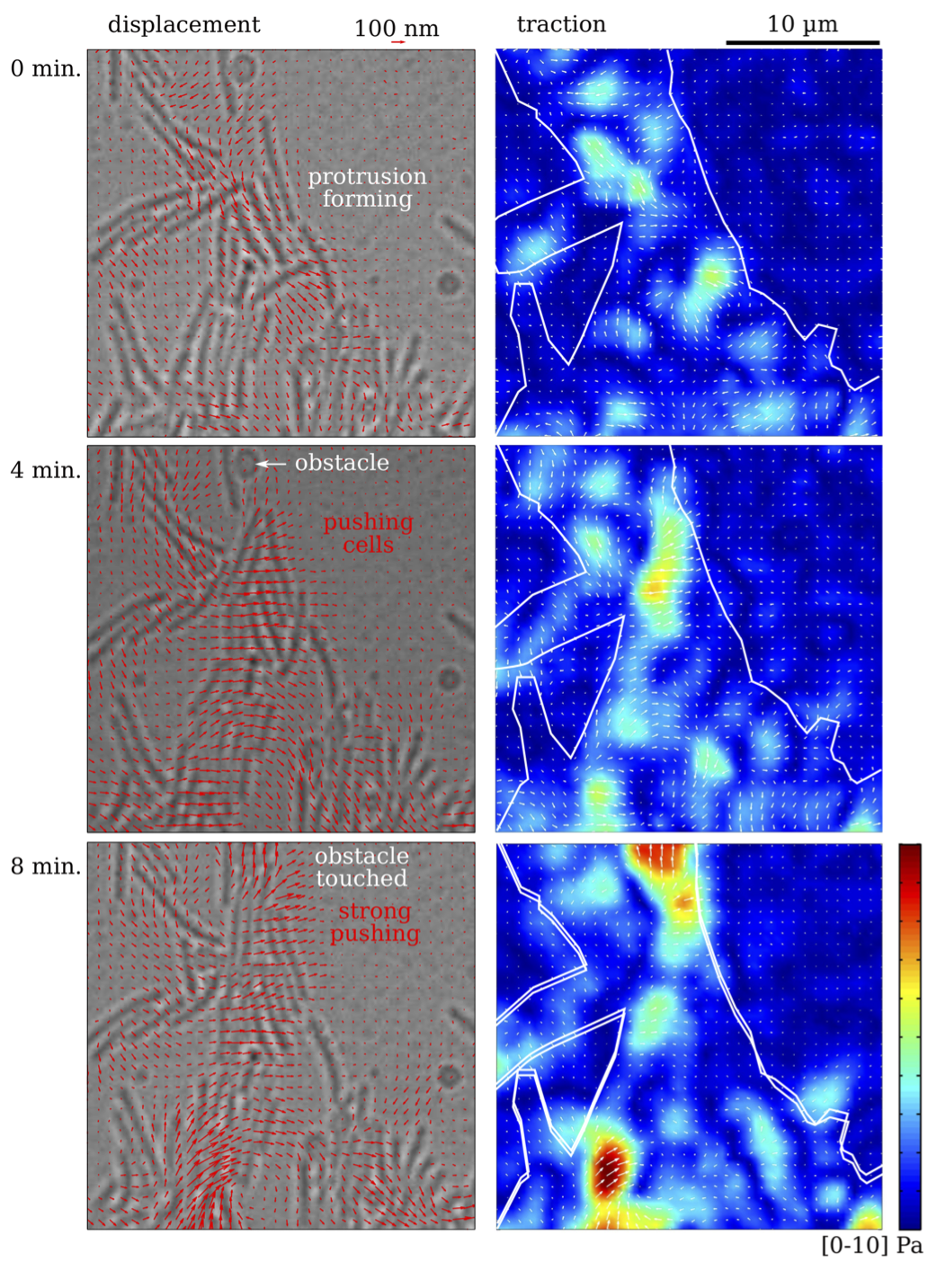} %
\caption{Gliding cell groups can form protrusions where cells collectively push in the direction of migration. 
At 8 minutes, pushing is amplified by contact with an obstacle. For better visibility, only every second 
vector is shown.} 
\label{fig_SI_gliding_groups}
\end{figure}

\begin{figure}[H]
  \centering
	\includegraphics[width=.9\linewidth]{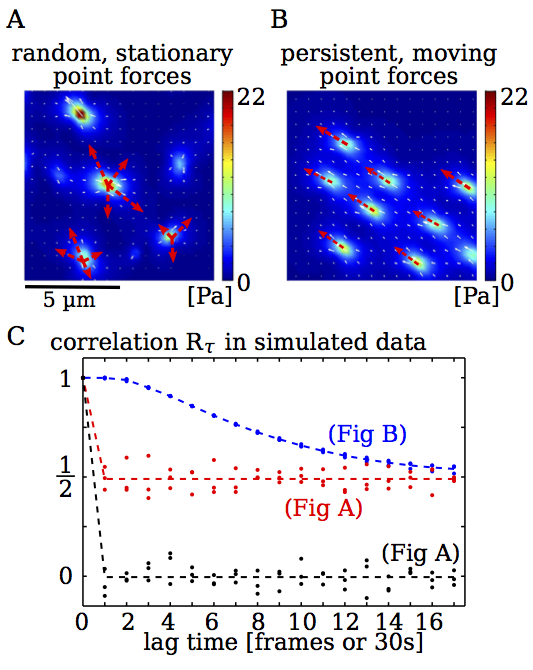} %
\caption{Simulation of artificial data to test the behavior of the correlation measure $R_{\tau}$. A)~Simulation of random stationary point forces. Forces are chosen in each frame from a Gaussian distribution. The resulting gel deformation is calculated for a vertical bead depth of $0.5\,\mum$ with a pixel size of $0.0607\,\mum$. Subsequently, the resulting displacement field is used as input for the calculation of traction and $R_{\tau}$. B)~Simulation of moving point forces of constant magnitude. Points move in the 
direction of force with a constant speed of $0.5\,\mum/\mathrm{s}$. 
C)~Correlation $R_{\tau}$ of reconstructed traction forces in simulations. Black dots: Simulation of stationary, random forces as shown in A). Gel displacements are calculated with respect to a stress-free reference state. As expected, we find $R_{\tau} \simeq 0$ for $\tau>0$. Red data: Simulation of stationary, random forces as shown in A), but displacements are now calculated with respect to the first frames of the movies. The pre-stress in the reference state leads to constant $R_{\tau} \simeq 0.5$ for finite time lag. Blue data: Simulations of persistently moving point forces as shown in B) where displacements are calculated with respect to the first frames of the movies. Motion of the forces leads to a slow decay of correlation towards $R_{\tau} \simeq 0.5$ for large lags.
For each condition, the data was generated from three simulations of each $35$ frames.} 
\label{fig_SI2}
\end{figure}

\bibliography{Mixo_biblio}

\end{document}